# Forecasting House Prices

Emanuel Kohlscheen*

Jan 2025

**Abstract**

This article identifies the factors that drove house prices in 13 advanced countries over the past 35 years. It does so based on Breiman's (2001) random forest model. Shapley values indicate that annual house price growth across countries is explained first and foremost by price momentum, initial valuations (proxied by price to rent ratios) and household credit growth. Partial effects of explanatory variables are also elicited and suggest important non-linearities, for instance as to what concerns the effects of CPI inflation on house price growth. The out-of-sample forecast test reveals that the random forest model delivers 44% lower house price variation root mean square errors (RMSEs) and 45% lower mean absolute forecast errors (MAEs) when compared to an OLS model that uses the same set of 10 pre-determined explanatory variables. Notably, the same model works well for all countries, as the random forest attributes minimal values to country fixed effects.

I would like to thank Pongpitch Amatyakul, Denis Gorea and Deniz Igan for very helpful comments. The views expressed in this paper are those of the author and do not necessarily reflect those of the Bank for International Settlements.

*Senior Economist. Bank for International Settlements, Centralbahnplatz 2, 4002 Basel BS, Switzerland. Email: emanuel.kohlscheen@bis.org .

# 1 Introduction

Houses are the main assets owned by households in most countries. Their purchases are typically the largest financial transaction that individuals carry out during their entire lifetime. And often their acquisition is also linked to the take up of their largest financial liability, in the form of mortgages. As a result, housing market dynamics are key for broader macroeconomic developments, with large effects on aggregate activity and household consumption patterns (Piazzesi and Schneider (2016); Duca et al (2021)). Further, the central role of credit in most acquisitions also makes the dynamics of house prices key for financial stability considerations.

In view of the above, and the difficulty to correctly predict house prices developments in advance, the present analysis tries to shed some new light on housing markets using a flexible non-parametric method from the machine learning literature – namely, a random forest model. The reason for a reassessment of these key markets with this tool was essentially given by Mullainathan and Spiess (2017), who noted that *"machine learning provides a powerful tool to hear, more clearly than ever, what the data have to say."*

Within machine learning, the current paper's focus on Breiman's (2001) random forest models follows from the results of Fernandez-Delgado et al (2014), who compared the performance of 179 classifier models across 121 datasets to find that random forests came out on top among all options.[1] Further, this method has the clear advantage that *"great out-of-sample performance [comes] without requiring subtle tuning"* (Athey and Imbens (2019), p. 696).

The main findings of the paper are the following. First, random forests clearly provide superior description of house price dynamics across advanced countries in-sample – as well as in pure forecasting exercises. This suggests that non-linearities are important. To give an idea of the magnitude of the improvement, relative to an OLS model that uses the same set of explanatory factors, the root square mean error (RMSE) and mean absolute error (MAE) reductions exceed 40%. Second, price momentum is the main factor explaining future house price developments, followed by initial valuations and credit growth. If anything, it finds that the last decade witnessed a relative increase in importance of population growth and of yield curves — the latter possibly because financial markets became more integrated globally, and housing became more attractive as an investment asset.

Third, partial effects from the random forest model suggest that house price growth reaches a maximum when inflation is in the 0–3% range, reaching a peak at 3%. Importantly, house price increases are found to fall short of inflation when CPI inflation surpasses 5%, and increasingly so when the pace of consumer price growth increases further. Put differently, housing is found to be a rather poor hedge against high levels of inflation.

Finally, the article finds that the same random forest model works well for all 13 advanced countries. When fed with the additional information about the country name in question, it largely ignores this information for almost every country. In all cases, the identification of the country is found to be less relevant even than the least important macro-financial variable. This may be taken as an indication that institutional differences in housing markets may be less relevant than previously suggested in the literature.

[1] Chakraborty and Joseph (2017) and Medeiros et al (2021) also found that random forests are best for prediction of general inflation.

**Relation to the literature.** The housing literature is very vast and reviewing it in its entirety would clearly be beyond the scope of this article. Instead, this section points out how the findings of the current study relate to selected seminal papers. Readers that desire more complete literature reviews, are referred to Leung (2004), Piazzesi and Schneider (2016) and Duca et al (2021).

First, the central finding that house price momentum is the single most important determinant of following year house price increases confirms Glaeser and Nathanson's (2017) conclusion that momentum at one-year horizon is a key determinant of house price dynamics internationally. As they note, this feature is typically missing in rational expectations models. Similarly, Granziera and Kozicki (2015) had found that house price dynamics in the United States were better explained by a model in which agents feature extrapolative rather than rational expectations. Relatedly, Ling et al (2015) stress the importance of sentiment for house prices. More generally, also Gu et al (2020) conclude that momentum or reversal tend to be the most powerful predictive signals for asset prices.

Second, the large relative importance of credit growth in driving house prices that is found here clearly resonates with earlier findings by Ėgert and Mihaljek (2007) and Mian and Sufi (2011, 2018), among others (see also the survey article by Duca et al (2021)). Further, the role of interest rates in driving house prices aligns with the conclusions of the seminal work by Muellbauer and Murphy (1997), who showed that the financial liberalization of mortgage markets in the United Kingdom greatly increased the significance of market interest rates in house price dynamics.

Third, while the finding that low CPI inflation maximizes house price inflation across all advanced countries is novel, it is noteworthy that Brunnermeier and Julliard (2008) already pointed out that a reduction in inflation leads to substantial house price increases when agents suffer money illusion. This is because they underestimate the cost of future mortgage payments when inflation falls, as they partly misinterpret this fall as a fall in the real interest rate. Importantly, the finding goes against the view that housing always benefits from higher inflation rates.

Speaking more broadly, the findings of this paper suggest that a very large fraction of aggregate house price movements can be attributed to variation in macroeconomic fundamentals, a result that was also in Glindro et al (2011), among others.

Overall, the novel contribution of the current article is to provide very precise data-driven quantifications of the economic effects that drive house price variations. It is shown that the random forest model leads to very large gains in house price forecasting performance. Further, the non-linear effects of CPI inflation on house prices are pinned down. Importantly, it is shown that the machine learning model is not a black-box, but that partial effects of the variables can be elicited. This powerful analytic tool can be added to the toolkit of policy makers, enhancing future analysis, and significantly diminishing forecast errors.

**Outline.** The article proceeds as follows. Section 2 presents the background and data on house price movements. The following Section presents and explains the models that are used to predict following year house price growth, as well as their in-sample performance. Section 4 discusses the test sample performance of the random forest model. Section 5 analyses the key drivers of house prices and Section 6 derives the associated partial effects. Section 7 shows how the overall inflation level affects house price pressures and Section 8 how the importance of key drivers has changed over time. Section 9 presents robustness checks, before some concluding remarks are offered.

# 2 Explaining House Prices in Advanced Economies

## 2.1 Yearly House Price Growth Across Countries and Time

The focus of the article is to examine the annual variation in nominal house price indices in advanced economies. For this, the study relies on the annual time series of these indices for 13 advanced economies between 1988 and 2019, i.e. a 31-year span. [2] [3] The source of these indices is the Bank for International Settlements. The country selection was based solely on data availability. This led to a cross-country sample consisting of Belgium, Canada, Denmark, Germany, Japan, Korea, Netherlands, New Zealand, Norway, Sweden, Switzerland, the United Kingdom and the United States. As a country that developed somewhat later, South Korea was included only from 2001 onwards. Cross-country house prices, as well as the full set of summary statistics used in the analysis that follows, can be seen in Figure A1 and Table A1 in the Appendix.

Figure 1 shows the kernel density distribution for nominal yearly house price growth for the 374 country-year observations that constitute the cross-country panel. On average, house price increased by 4.3% per year, with a standard deviation of 6.2%, a neutral skewness of 0.03 and a kurtosis of 4.4. Overall, the mean variation compares favourably to the mean CPI inflation rate of 2.1%. Interestingly, the high kurtosis indicates that house price variations are leptokurtic, i.e. they have a light tailed historical distribution. In other words, very large variations are not very common.

[2] The analysis focuses on medium-term variations, i.e. at annual frequency. While quarterly price series are available in many cases, very short-term fluctuations are much more volatile due to a variety of factors, including seasonal variations.

[3] The reason for the start date is the length of the time series for the explanatory variables presented later. The main constraining factor is information on the long end of the yield curves.

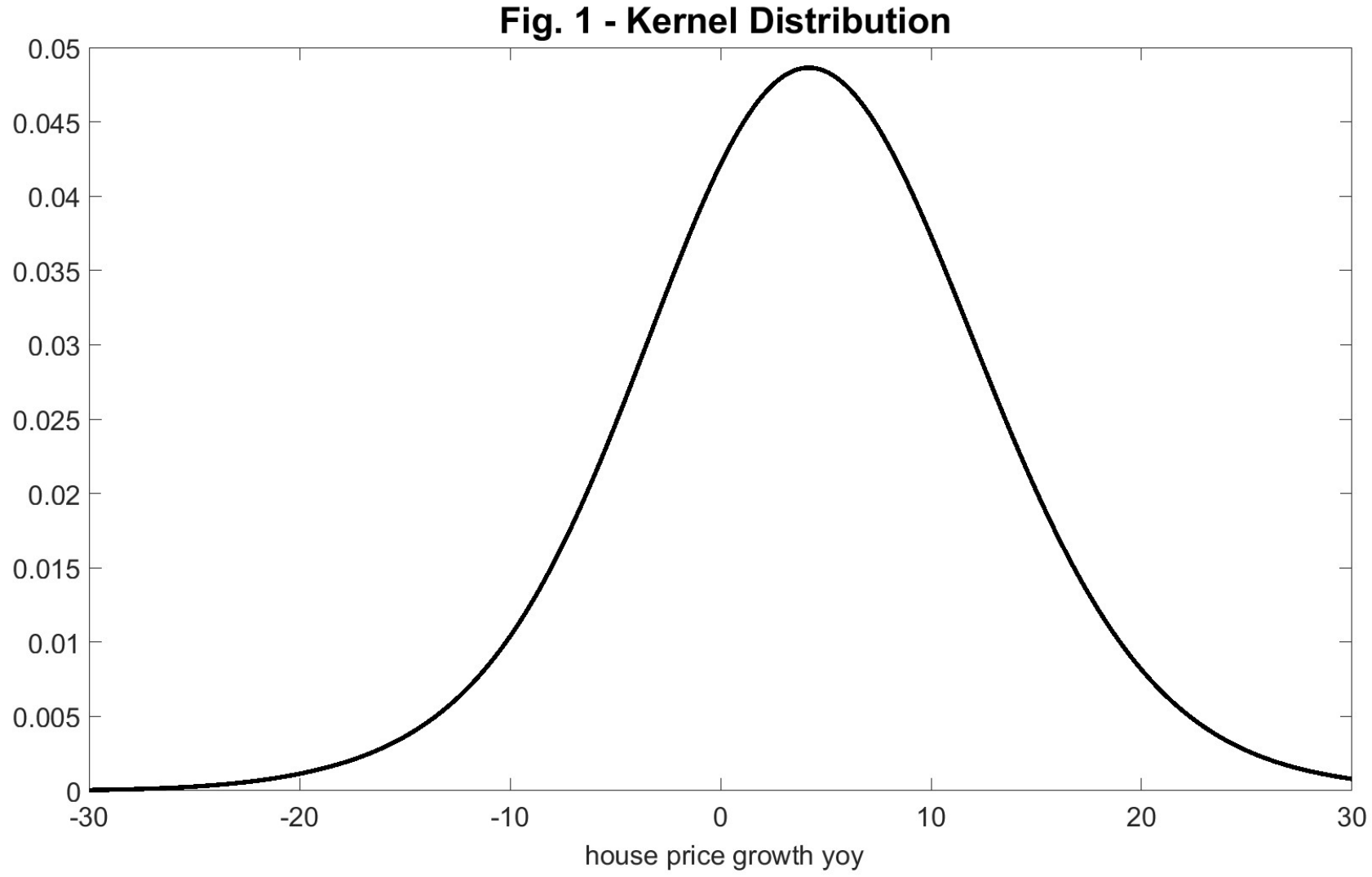

Kernel density function of yearly house price increases, in per cent, for the 374 country-year observations of the panel. Average house price growth in sample is 4.3% per year, with a standard deviation of 6.2

## 2.2 Potential Explanatory Factors

House price dynamics are affected by market psychology (i.e. whether the market is in "boom" or "bust" mode), current price levels (valuations), macroeconomic, financial and demographic factors. The main potential explanatory factors that were used to explain house price growth in this study follow below. These are then included throughout in the baseline econometric and machine learning models.

The 1-year forward nominal house price growth can in principle be affected by current house price momentum (see e.g. Glaeser and Nathanson (2017)). This could be a reflection of the psychology that is prevailing in markets. Buyers and sellers might for instance extrapolate future price movements from past ones. Thus, any model of housing should allow for the possibility of an inertial price movement component. This is the AR (1) element included in all models of the current paper.

Current or initial valuations can also affect future price dynamics. This variable captures the effect that over- or undervaluation can have on future price dynamics. More specifically, in the current analysis, the valuation metric is the aggregate price to rent ratio index that is published regularly by the OECD.

House prices are also potentially affected by activity and consumer price pressure variables, that is, GDP growth which captures the evolution of income and CPI inflation. While the prior is that GDP growth boosts house price growth, by increasing agents' capability to purchase houses (Ortalo-Magné and Rady (2006)), CPI inflation can have an ambiguous effect. This is because higher inflation raises housing construction and maintenance costs. It may also boost prices if housing is perceived as a good hedge against inflation (Tsatsaronis and Zhu (2004)). That said, it could also lead to lower demand for housing if agents suffer from money illusion, as has been posited by Brunnermeier and Julliard (2008).

Construction and housing are typically finance intensive. Therefore, financial market variables and credit are key for its pricing. Thus, also nominal household credit growth is included as one potential driver. Further, the short- as well as the long-term interest rate (obtained from the OECD), as well as the stock market gain and an indicator of market volatility (i.e. the VXO index[4]) are considered as potential explanatory factors.

Finally, demographic factors could be critical for housing demand. To account for this, population growth is considered as the tenth and last potential explanatory variable. The precise definition of all variables, as well as their sources are listed in the Appendix.[5]

## 3 Model selection

Three alternative types of models are considered for explaining house price growth. All of them rest exclusively on pre-determined variables, so that the model can always be interpreted as predictive. In the current Section though, this refers to in-sample prediction.

The first two models are the econometric benchmarks: the naïve AR (1) model, which tries to explain 1-year forward house price growth with current house price growth only; and the linear least squares model, which explains it with the full list of pre-determined potential explanatory factors listed in the previous section.

The third model is a machine learning model. More specifically, the random forest model developed by Breiman (2001). Random forests are an ensemble of regression trees.[6] As pointed out in the introduction, the main advantages of this non-parametric method are that it is able to capture non-linearities and complex interactions between the variables, at the same time that it is relatively transparent in that it does not require subtle tuning of many hidden hyperparameters (Athey and Imbens (2019)). Further, its strong forecasting performance has been attested, for instance, by Fernandez-Delgado et al (2014) and Medeiros et al (2021).

A regression tree is essentially a simple classification algorithm that follows the sequence

I) randomly select 2/3 of the observations for the training sample;

II) start with a node containing the training sample observations, and for each possible explanatory variable ("feature") and each possible associated threshold level, compute the sum of mean squared prediction errors that would result after the split. The predictions are the average outcomes of 1-year forward house price growth for all observations in the two resulting nodes;

[4]This is an index that is similar in spirit to the better known VIX implicit equity market volatility index. Its advantage is that the time series is longer.

[5]The complete dataset as well as the code used for this study can also be obtained upon request.

[6]Regression trees are explained in Breiman et al (1984).

III) split the observations at the node based on the variable and associated threshold that delivers the lowest sum of mean square prediction errors after the split. That is, the choice of predictor $k$ and an associated threshold $t_k$ at which the split happens, is selected according to the expression

$$argmin(k, t_k) = \frac{S_A.MSE(S_A) + S_B.MSE(S_B)}{S},$$

where S corresponds to the number of training points, and A and B represent the subsets of the training sample after the binary split.

IV) For each resulting node, return to step I).

The loop continues until each final node has no more observations in it than what the stopping criterion prescribes. Simultaneously, to inject randomness, only a random subset of explanatory variables is used at each split.

Figure A2 in the Appendix illustrates with the first created regression tree that this flexible but highly disciplined computational algorithm delivers. To attain greater robustness, the purely data-driven method grows not only one regression tree, but a large number of them, and then takes the average of the outcomes over all the trees of the forest.

Mean square error of house price growth forecasts as a function of the number of trees in the random forest model (between 1 and 200). For each tree the minimum number of observations per parent node is held fixed at 10.

Figure 2 shows how the mean square error of the cross-country house price growth model evolves as the number of trees in the random forest grows. The errors drop sharply after a few more regression trees are included in the ensemble. After 100 trees are included, any further gains in fitting performance become rather marginal.

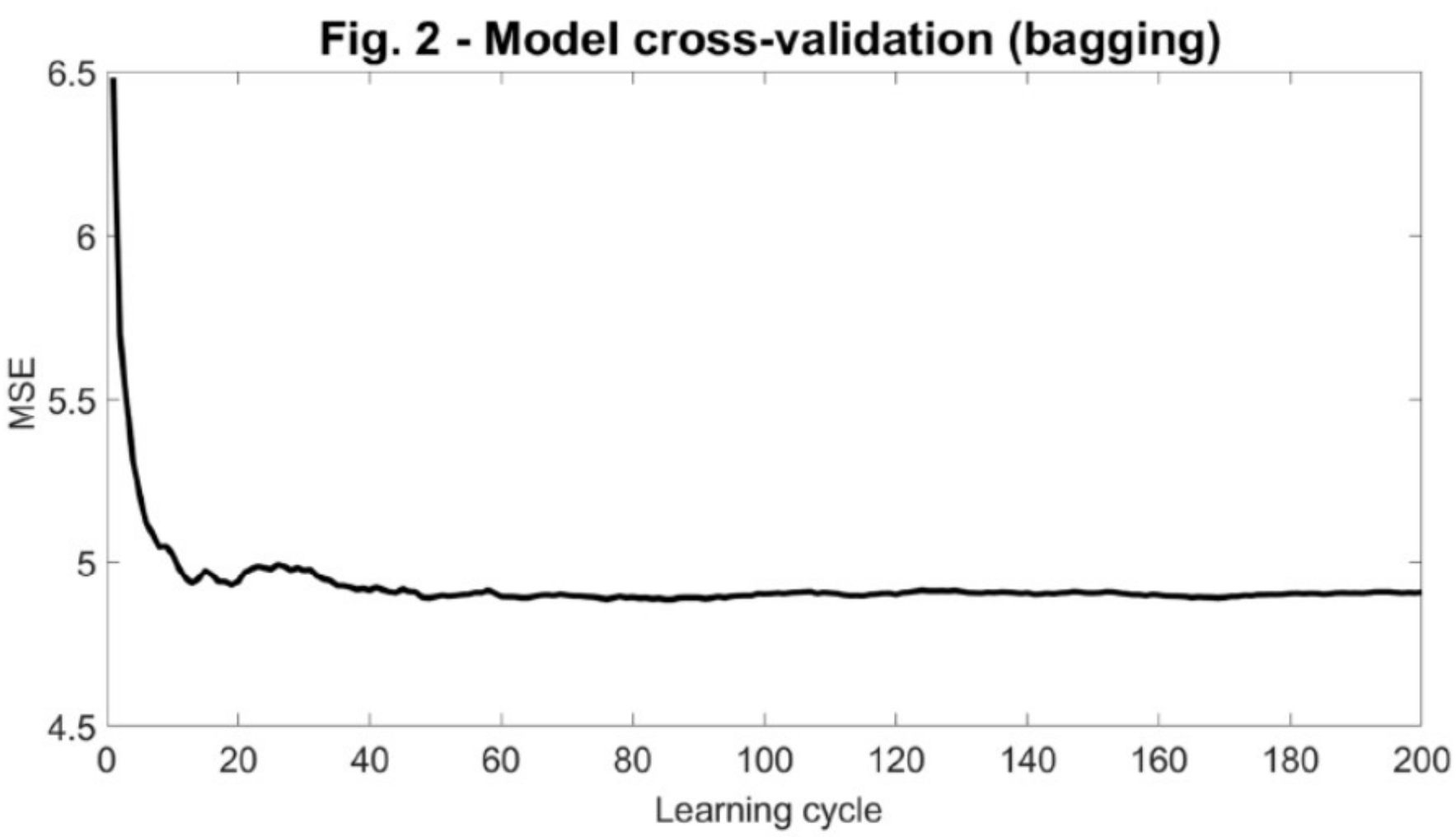

Mean square error of house price growth forecasts as a function of the number of trees in the random forest model (between 1 and 200). For each tree the minimum number of observations per parent node is held fixed at 10.

The results of a much more complete examination of the forecasting performance of the 10-variable house price model are presented in Table 1. Here, the resulting RMSEs and MAEs are shown for the naïve AR(1), the OLS and the random forest models in-sample.

First, the AR (1) model is able to explain 35.3% of the variance in 1-year forward house prices (adjusted R2 statistic), while the 10-variable OLS model accounts for 44.0% of the variation. As to what concerns RMSEs, that of OLS is 7.0% below of the AR(1) model in-sample. In turn, the mean absolute error of the OLS model is 8.8% below the one of the minimalistic autoregressive benchmark.

In turn, the random forest model outperforms the AR (1) and the OLS model by a wide margin, irrespective of the tree growth stopping criterion (i.e., the maximum number of observations per terminal node of the tree) and/or the number of trees that is used in the ensemble. All RMSE ratios are well below unity, indicating a better in-sample data description.

One could think that the better performance of random forests is due to in-sample overfitting. Indeed, the Table shows that performance is better when very deep trees are grown (such as max obs / terminal node = 5), than when very shallow trees are grown (max obs / terminal node = 30). Further, performance tends to increase marginally when a larger number of trees is utilized in the ensemble (as was already indicated by Figure 2).

**Table 1 – Predictive house price model: Root mean square errors and mean absolute errors**

| | | Full sample (in sample) | | | | | | | | | |
|---|---|---|---|---|---|---|---|---|---|---|---|
| max obs / terminal node | number of trees | RMSE | | | MAE | | | ML RMSE ratio to | | MAE ratio to | |
| | | AR(1) | OLS | RF | AR(1) | OLS | RF | AR(1) | OLS | AR(1) | OLS |
| | | 4.980 | 4.630 | | 3.900 | 3.555 | | | | | |
| 5 | 50 | | | 1.971 | | | 1.397 | 0.396 | 0.426 | 0.358 | 0.393 |
| | 100 | | | 1.953 | | | 1.366 | 0.392 | 0.422 | 0.350 | 0.384 |
| | 200 | | | 1.950 | | | 1.362 | 0.392 | 0.421 | 0.349 | 0.383 |
| | 500 | | | 1.947 | | | 1.359 | 0.391 | 0.420 | 0.349 | 0.382 |
| 10 | 50 | | | 2.519 | | | 1.727 | 0.506 | 0.544 | 0.443 | 0.486 |
| | 100 | | | 2.446 | | | 1.687 | 0.491 | 0.528 | 0.433 | 0.475 |
| | 200 | | | 2.391 | | | 1.672 | 0.480 | 0.516 | 0.429 | 0.470 |
| | 500 | | | 2.380 | | | 1.661 | 0.478 | 0.514 | 0.426 | 0.467 |
| 15 | 50 | | | 2.725 | | | 1.923 | 0.547 | 0.589 | 0.493 | 0.541 |
| | 100 | | | 2.747 | | | 1.947 | 0.552 | 0.593 | 0.499 | 0.548 |
| | 200 | | | 2.706 | | | 1.912 | 0.543 | 0.584 | 0.490 | 0.538 |
| | 500 | | | 2.730 | | | 1.913 | 0.548 | 0.590 | 0.491 | 0.538 |
| 20 | 50 | | | 3.014 | | | 2.134 | 0.605 | 0.651 | 0.547 | 0.600 |
| | 100 | | | 2.986 | | | 2.137 | 0.599 | 0.645 | 0.548 | 0.601 |
| | 200 | | | 2.976 | | | 2.124 | 0.598 | 0.643 | 0.545 | 0.597 |
| | 500 | | | 2.957 | | | 2.104 | 0.594 | 0.639 | 0.540 | 0.592 |
| 25 | 50 | | | 3.194 | | | 2.256 | 0.641 | 0.690 | 0.579 | 0.635 |
| | 100 | | | 3.158 | | | 2.260 | 0.634 | 0.682 | 0.579 | 0.636 |
| | 200 | | | 3.157 | | | 2.257 | 0.634 | 0.682 | 0.579 | 0.635 |
| | 500 | | | 3.165 | | | 2.266 | 0.636 | 0.684 | 0.581 | 0.637 |
| 30 | 50 | | | 3.357 | | | 2.435 | 0.674 | 0.725 | 0.624 | 0.685 |
| | 100 | | | 3.339 | | | 2.395 | 0.670 | 0.721 | 0.614 | 0.674 |
| | 200 | | | 3.325 | | | 2.389 | 0.668 | 0.718 | 0.613 | 0.672 |
| | 500 | | | 3.321 | | | 2.384 | 0.667 | 0.717 | 0.611 | 0.671 |

Note: Ratios below indicate that ML outperforms AR(1) or OLS models.

If one takes a stopping criterion of 10, and random forests of 500 trees, the reduction in RMSE of the house price model is of 48.6% relative to the OLS model that uses the very same set of explanatory factors. The MAE reduction is of 53.3%.

The above demonstrates that the random forest provides a much better data description in-sample, as had already been indicated by Mullainathan and Spiess (2017). The large gains likely reflect the fact that the linear functional form ingrained in OLS is simply too restrictive. At the same time, random forests have the capability to capture more complex interactions between variables.

## 4 Post-Estimation Sample Prediction Performance

The natural question that follows is if superior performance carries over to a post-estimation sample. To analyze this, the three models are re-estimated in this Section using only information between 1988 and 2014. The years 2015-2019 are left for out-of-sample evaluation.

Table 2 presents the performance statistics for this test period. What is clear is that forecasting performance in the post-estimation is only moderately inferior than in-sample. Again, all RMSE ratios are well below 1.0. Taking the benchmark (10 obs/terminal node, 500 trees) specification again, shows that even out-of-sample, the RMSE reduction in the random forest is of no less than 50.3% relative to the AR(1) model, and of 44.0% relative to the OLS model that uses the very same explanatory variables. The corresponding MAE reductions are an impressive 49.0% and 44.6%.

**Table 2 – Predictive house price model: Post-Estimation Sample**

| | | Test sample performance (2015-2019) | | | | | | | | | |
|---|---|---|---|---|---|---|---|---|---|---|---|
| max obs / | number of | RMSE | | | MAE | | | ML RMSE ratio to | | MAE ratio to | |
| terminal node | trees | AR(1) | OLS | RF | AR(1) | OLS | RF | AR(1) | OLS | AR(1) | OLS |
| | | 5.461 | 4.847 | | 3.892 | 3.588 | | | | | |
| 5 | 50 | | | 2.432 | | | 1.739 | 0.445 | 0.502 | 0.447 | 0.485 |
| | 100 | | | 2.428 | | | 1.759 | 0.444 | 0.501 | 0.452 | 0.490 |
| | 200 | | | 2.312 | | | 1.693 | 0.423 | 0.477 | 0.435 | 0.472 |
| | 500 | | | 2.291 | | | 1.699 | 0.419 | 0.473 | 0.437 | 0.474 |
| 10 | 50 | | | 3.049 | | | 2.142 | 0.558 | 0.629 | 0.550 | 0.597 |
| | 100 | | | 2.667 | | | 1.994 | 0.488 | 0.550 | 0.512 | 0.556 |
| | 200 | | | 2.800 | | | 2.069 | 0.513 | 0.578 | 0.532 | 0.577 |
| | 500 | | | 2.715 | | | 1.987 | 0.497 | 0.560 | 0.510 | 0.554 |
| 15 | 50 | | | 2.952 | | | 2.182 | 0.541 | 0.609 | 0.560 | 0.608 |
| | 100 | | | 3.080 | | | 2.299 | 0.564 | 0.635 | 0.591 | 0.641 |
| | 200 | | | 3.179 | | | 2.311 | 0.582 | 0.656 | 0.594 | 0.644 |
| | 500 | | | 3.144 | | | 2.300 | 0.576 | 0.649 | 0.591 | 0.641 |
| 20 | 50 | | | 3.425 | | | 2.540 | 0.627 | 0.707 | 0.653 | 0.708 |
| | 100 | | | 3.430 | | | 2.558 | 0.628 | 0.708 | 0.657 | 0.713 |
| | 200 | | | 3.398 | | | 2.483 | 0.622 | 0.701 | 0.638 | 0.692 |
| | 500 | | | 3.380 | | | 2.487 | 0.619 | 0.697 | 0.639 | 0.693 |
| 25 | 50 | | | 3.724 | | | 2.775 | 0.682 | 0.768 | 0.713 | 0.773 |
| | 100 | | | 3.528 | | | 2.638 | 0.646 | 0.728 | 0.678 | 0.735 |
| | 200 | | | 3.496 | | | 2.576 | 0.640 | 0.721 | 0.662 | 0.718 |
| | 500 | | | 3.645 | | | 2.707 | 0.667 | 0.752 | 0.696 | 0.755 |
| 30 | 50 | | | 3.737 | | | 2.807 | 0.684 | 0.771 | 0.721 | 0.782 |
| | 100 | | | 3.895 | | | 2.820 | 0.713 | 0.803 | 0.724 | 0.786 |
| | 200 | | | 3.776 | | | 2.833 | 0.691 | 0.779 | 0.728 | 0.790 |
| | 500 | | | 3.785 | | | 2.818 | 0.693 | 0.781 | 0.724 | 0.785 |

Note: Ratios below indicate that ML outperforms AR(1) or OLS models.

The modest decrease in performance relative to the in-sample case, is partly related to Breiman's (2001) observation that random forests do not seem to overfit - because they rely heavily on a law of large numbers. Even so, to err on the side of caution, the (10, 500) model is taken as the benchmark in what follows. That is, somewhat shallower trees are used, even though the (5, 500) model would deliver even lower RMSEs. As it turns out, this conservative choice does not affect the broad pattern of results.

## 5 Shapley Values

One can analyse the relative importance of the different drivers of house prices across countries and time by looking at the mean absolute values of the Shapley values. For each observation, the Shapley value indicates by how much the predicted outcome (house price growth) varies when the given variable is added to the set. Taking the average of the absolute values of these across all observations and variables then gives the predictor importance statistic.

Figure 3 shows that for the 1988-2019 period, momentum was the key factor explaining house price movements. At least in the medium-term – i.e. at yearly frequency – price house growth displays some hysteresis. This provides international confirmation to the finding of Glaeser and Nathanson (2017). It likely captures whether the market is in a boom or bust mode that so often characterizes housing markets for protracted periods.

The large importance of momentum is of course also entirely consistent with a slow pace of adjustment or corrections of disequilibria. To wit, Alvarez-Román and García-Posada (2021), for instance, find that only 8–15% of the deviations from equilibrium prices are removed within one year in the case of Spain. Burnside et al (2016) rationalize that protracted housing boom-bust cycles are generated in a model with heterogeneous expectations about long-run fundamentals in which agents change their views over time because of social interactions.

Valuation, as captured by the price to rent ratio, is the second most important factor, followed closely by household credit growth. Also, GDP growth and inflation clearly impact the market,[7] as do interest rates and population growth. Last, equity market developments and implied volatilities are the least important factors, even though no time dummies were included in the specification. The latter suggests that housing market dynamics do not follow the same logic as equity markets.

[7] As for what concerns the effect of output, note that Ortalo-Magné and Rady (2006) already pointed out that changes in income variations can have a large effect on house prices because of its effects on the ability of young households to afford down payments on their first home.

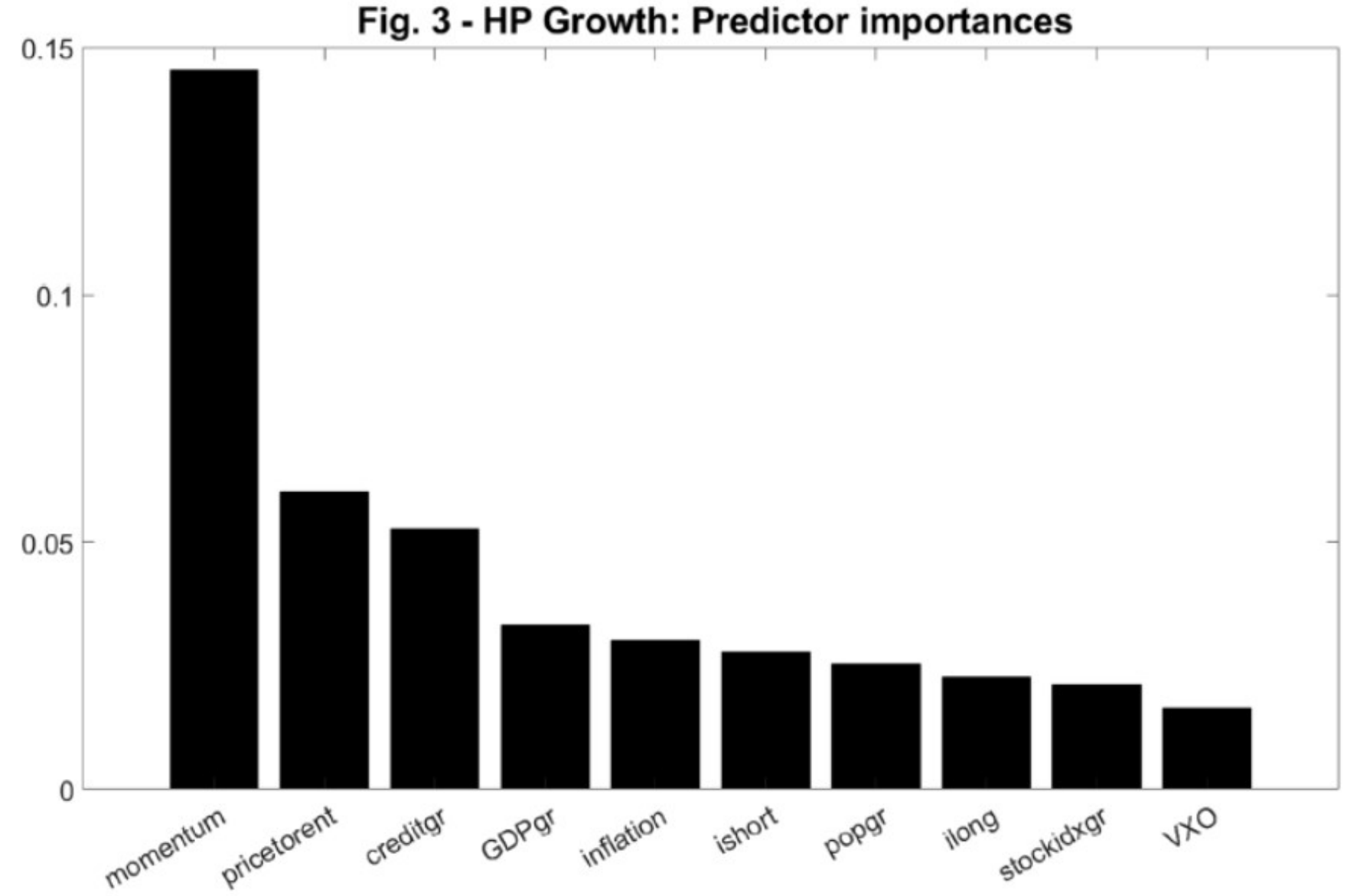

**Mean absolute Shapley values (predictor importances) for each of the 10 features used in the baseline house price growth model.**

## 6 Eliciting Partial Effects

The analysis of the drivers of house prices can be taken one step further: partial effects of (pre-determined) explanatory variables can be elicited. This can be done by varying the covariate of interest, while maintaining other covariates fixed at their sample mean – just as one would do for partial effects in econometrics. The main difference is that here the effects do not need to be linear.

In what follows, the analysis focuses on the variables that are directly impacted by monetary policy, as well as the main economic contributor to house price dynamics according to the Shapley values. Figure 4 shows these partial effects along the price to rent ratio and the policy interest rate. That is, it projects the house price variation that is predicted by the random forest model when these two dimensions vary, while the eight remaining variables are held fixed at their sample means.

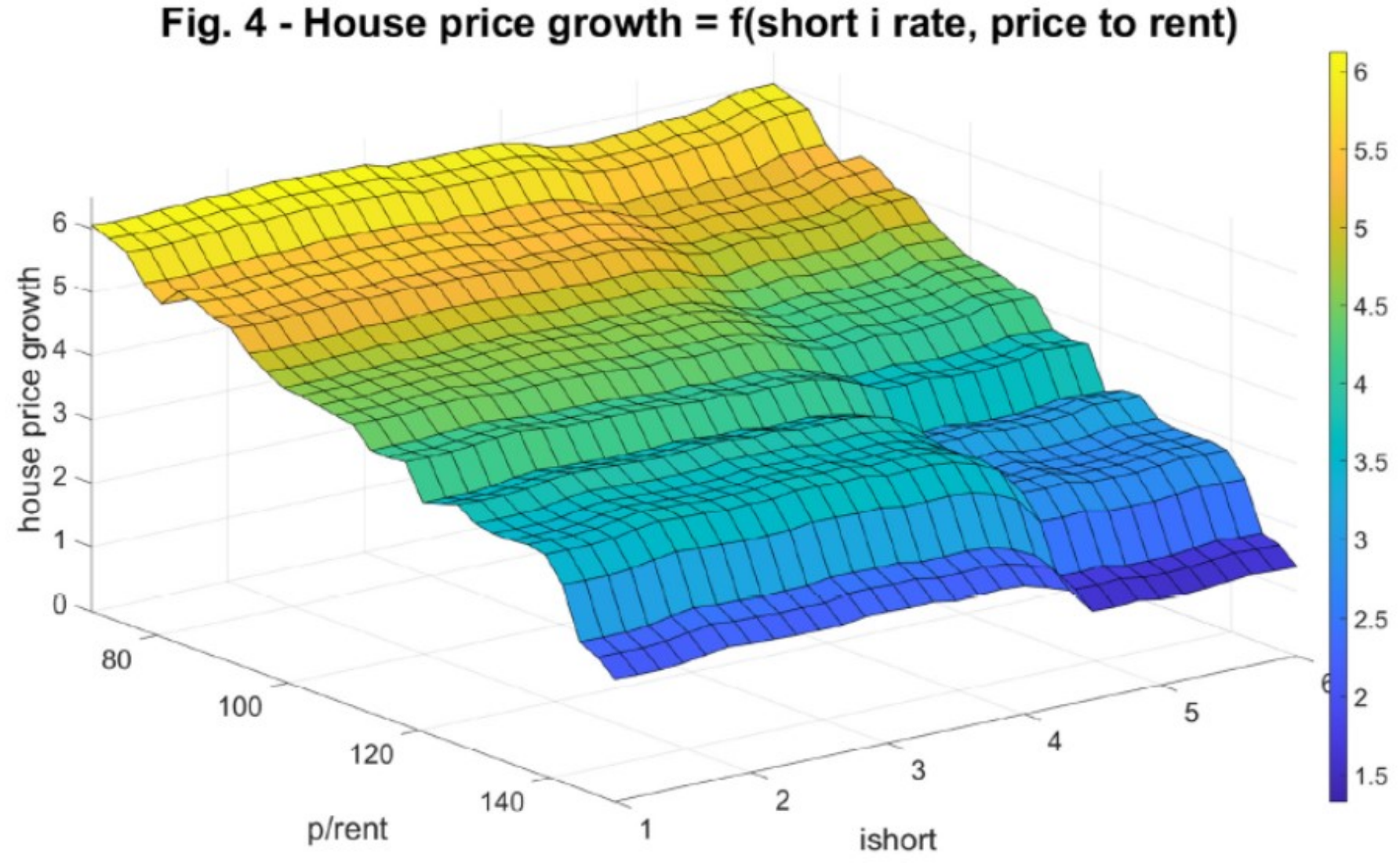

Fitted values for the baseline RF house price growth model as a function of the price to rent ratios and short-term interest rates. The remaining eight features in the model are fixed at their sample means.

The fitted house price increase rises to 6% per year when the price to rent ratio in the preceding year is low, but falls to below 3% when initial valuations are already high. In turn, short-term interest rates on their own push predicted house price growth down modestly, with effects becoming slightly stronger as rates move above 4%.

# 7 Effects of CPI Inflation on House Prices and an Example

One can also tease out the effect that general inflation has on house prices. This is shown on Figure 5, which is based on an exercise that is similar to the one in the previous section. All else equal, house price growth is highest when inflation is in the 0–3% range, reaching a peak at 3%. Once CPI inflation surpasses this level, it starts to have a dampening effect on house price growth. Importantly, this pattern is robust to alternative initial valuations.

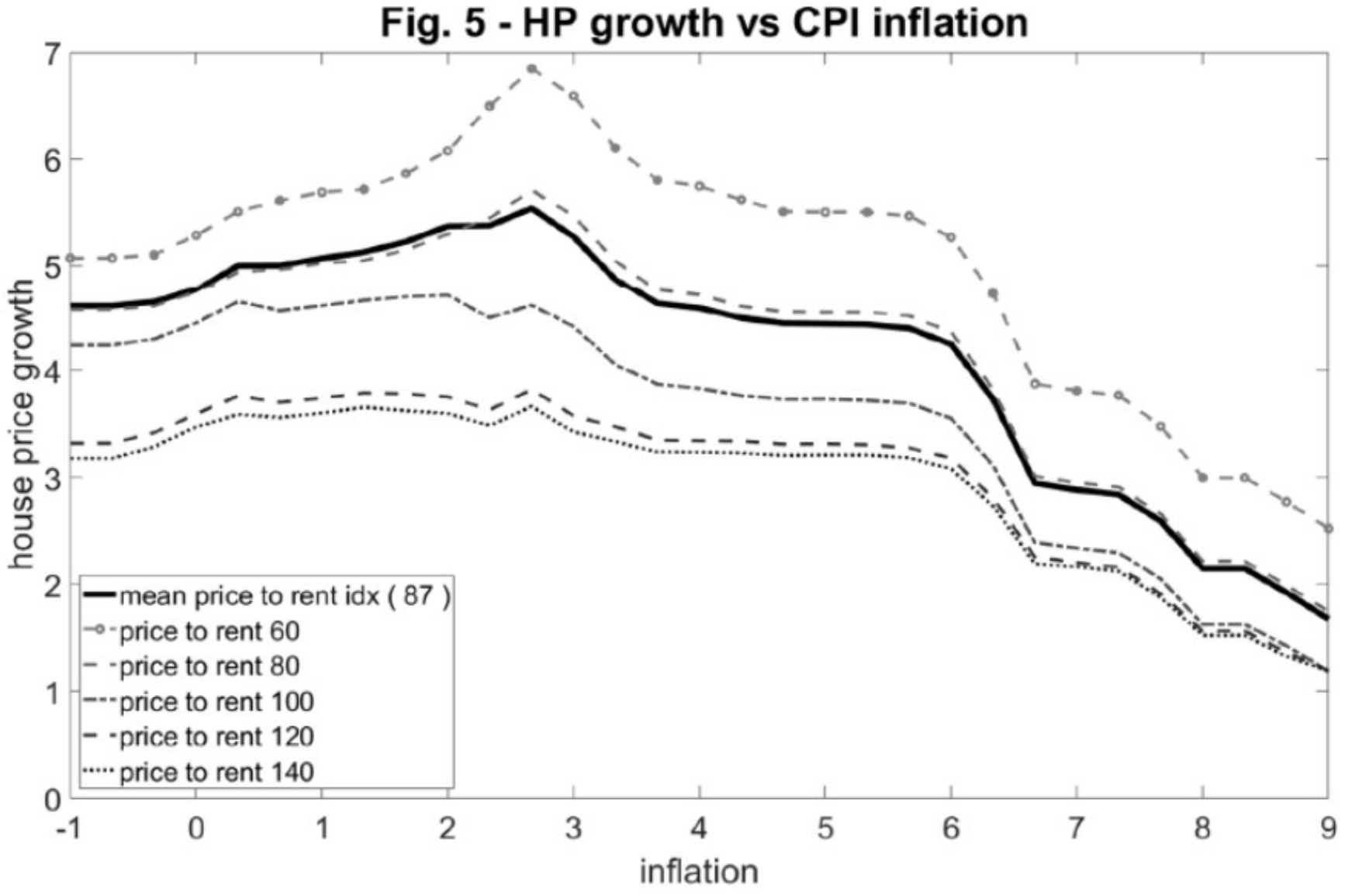

**Fitted values for the baseline RF house price growth model as a function of CPI inflation for different values of the price to rent ratio. The remaining eight features in the model are fixed at their sample means.**

As a purely illustrative example, Figure 6 shows how the random forest model would forecast 2019 house prices, based on 2018 data (the horizontal axis). This is compared with actual outcomes across countries (the vertical axis). Note that these predictions are based on models that are estimated on data to 2018 only (the results of the more systematic analysis where already presented in Section 4).

Overall, the RF model explains about 43% of the cross-country variation in the last year of the sample. Figure 7 shows that the OLS model on the same pre-determined variables is able to explain only 19% of the cross-country variation.

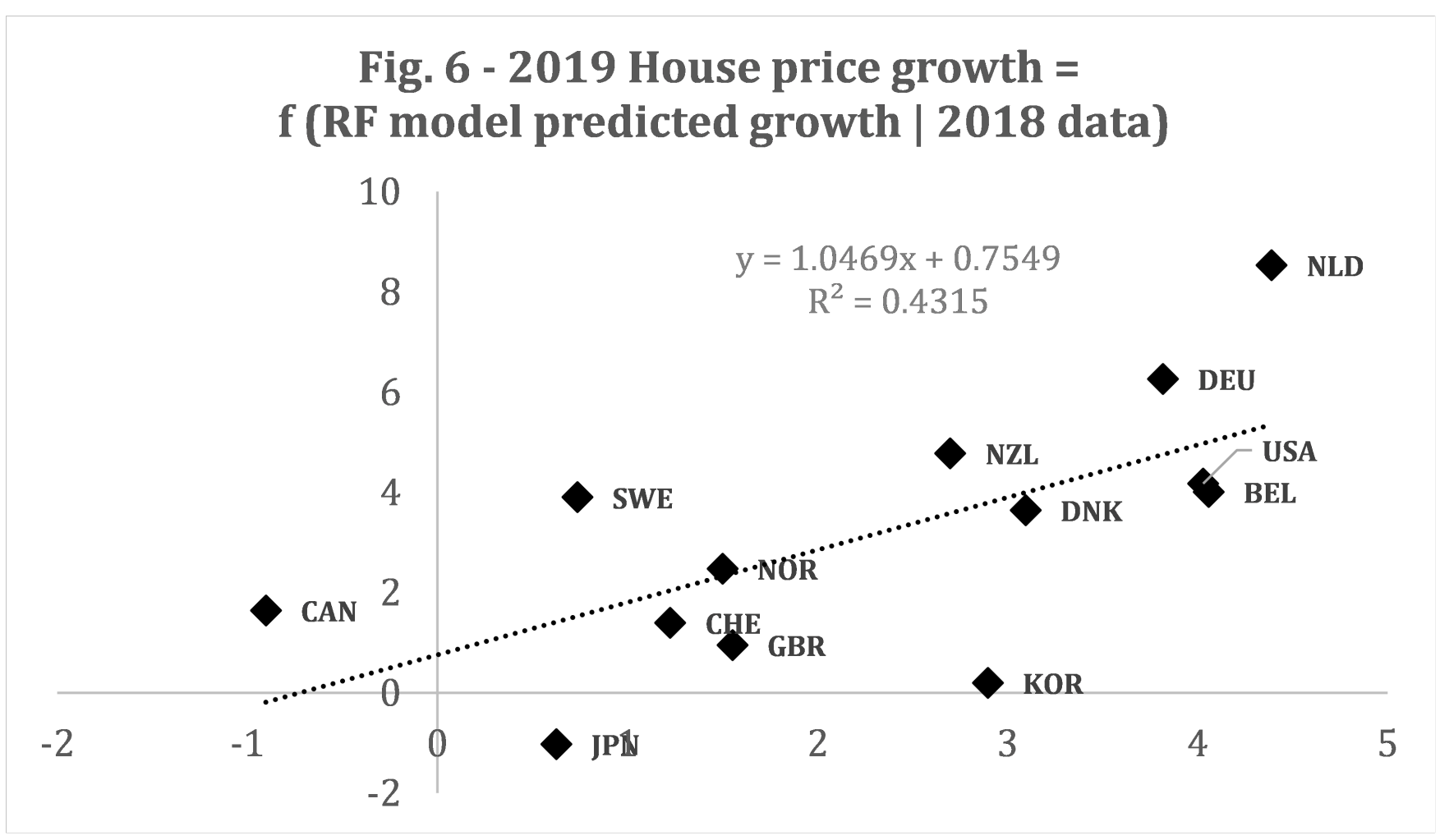

Note: This graph shows actual nominal house price growth in 2019 by country in % (vertical axis), against the house price growth predicted by the random forest model (horizontal axis). The random forest model that was trained using data between 1988 and 2018. Predicted values for 2019 are obtained by feeding this fixed model with 2018 data.

**Fig. 7 - 2019 House price growth = f (OLS model predicted growth | 2018 data)**

y = 0.8355x + 0.5895
$R^2$ = 0.1942

Note: This graph shows actual nominal house price growth in 2019 by country in % (vertical axis), against the house price growth predicted by an OLS model (horizontal axis). The OLS model was estimated on data between 1988 and 2018. Predicted values for 2019 are obtained by feeding this fixed model with 2018 data.

## 8 Time Variation

Have the drivers of house price changes varied across time? Time variations could result for instance from a gradual liberalization of mortgage markets, or from increasing financial globalization - which could lead to greater co-movement of yield curves and credit across countries.

Table 3 shows how the average of the absolute Shapley values varies by sub-sample. Here, predictor importances have been normalized so as to add up to 1 in each period. If anything, the weight of initial valuations and credit growth has decreased on the margin during the last decade. That said, momentum and valuation continued to be the main drivers of house price changes.

Yield curve and population growth have become somewhat more important as driving forces in the more recent period. These findings would appear to align well with the observation that - rather than being driven by credit booms - housing became more attractive as an investment instrument during this period (see e.g. Igan et al (2022) on this). Additionally, major central banks engaged in quantitative easing policies during the past decade. Part of this added liquidity likely found its way to asset markets, including houses.

**Table 3 - Normalized Predictor Importance by Period**

| | 1988-2019 | 2011-2019 |
|---|---|---|
| momentum | 0.330 | 0.277 |
| price / rent | 0.140 | 0.097 |
| credit growth | 0.121 | 0.062 |
| GDP growth | 0.074 | 0.072 |
| CPI inflation | 0.069 | 0.069 |
| i short | 0.067 | 0.087 |
| pop. growth | 0.059 | 0.090 |
| i long | 0.054 | 0.073 |
| stock return | 0.048 | 0.141 |
| VXO | 0.038 | 0.032 |

# 9 Robustness

Even though random forests rely to a large extent on the law of large numbers, and the specification options were already discussed in Section 3, a number of additional robustness checks were performed. Two of these are shown below.

First, an alternative model was tested in which a full set of country dummies was added to the set of explanatory variables in all models. The predictor importance factors that resulted (which are shown in Figure 8), indicate that – perhaps a bit surprisingly – the RF model systematically regarded the country identification as the least important information. If any, the most relevant country dummy was that for Japan – which nevertheless was less important than the market volatility index VXO. One way to interpret this finding is that cross-country heterogeneity is likely to already be captured by the relatively comprehensive list of explanatory variables.

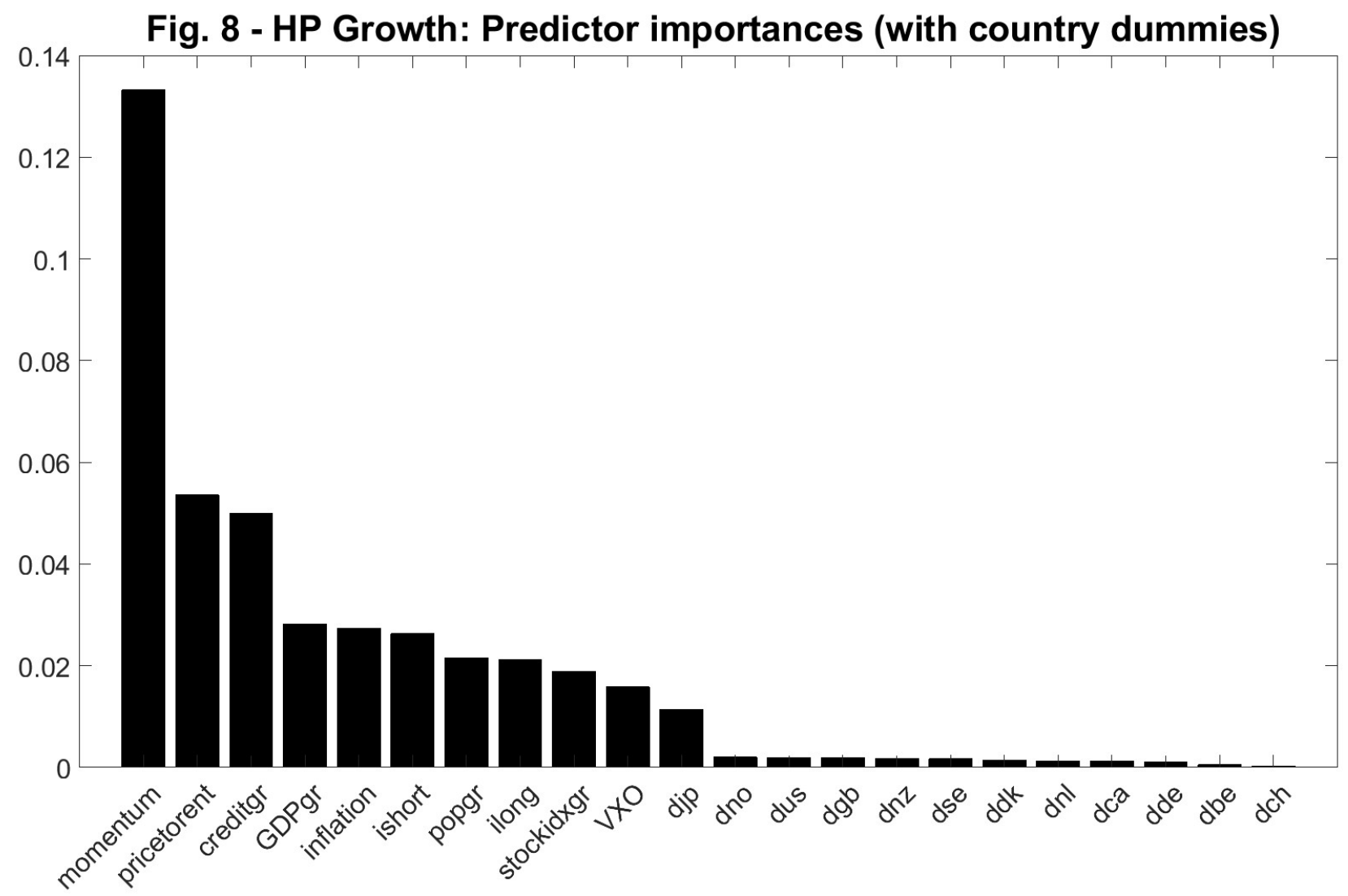

Second, Figure 9 shows predictor importances when the stopping criterion in the RF is set at 5 observations per terminal node, instead of 10. Overall, this leads only to very minor relative changes in predictor importance.

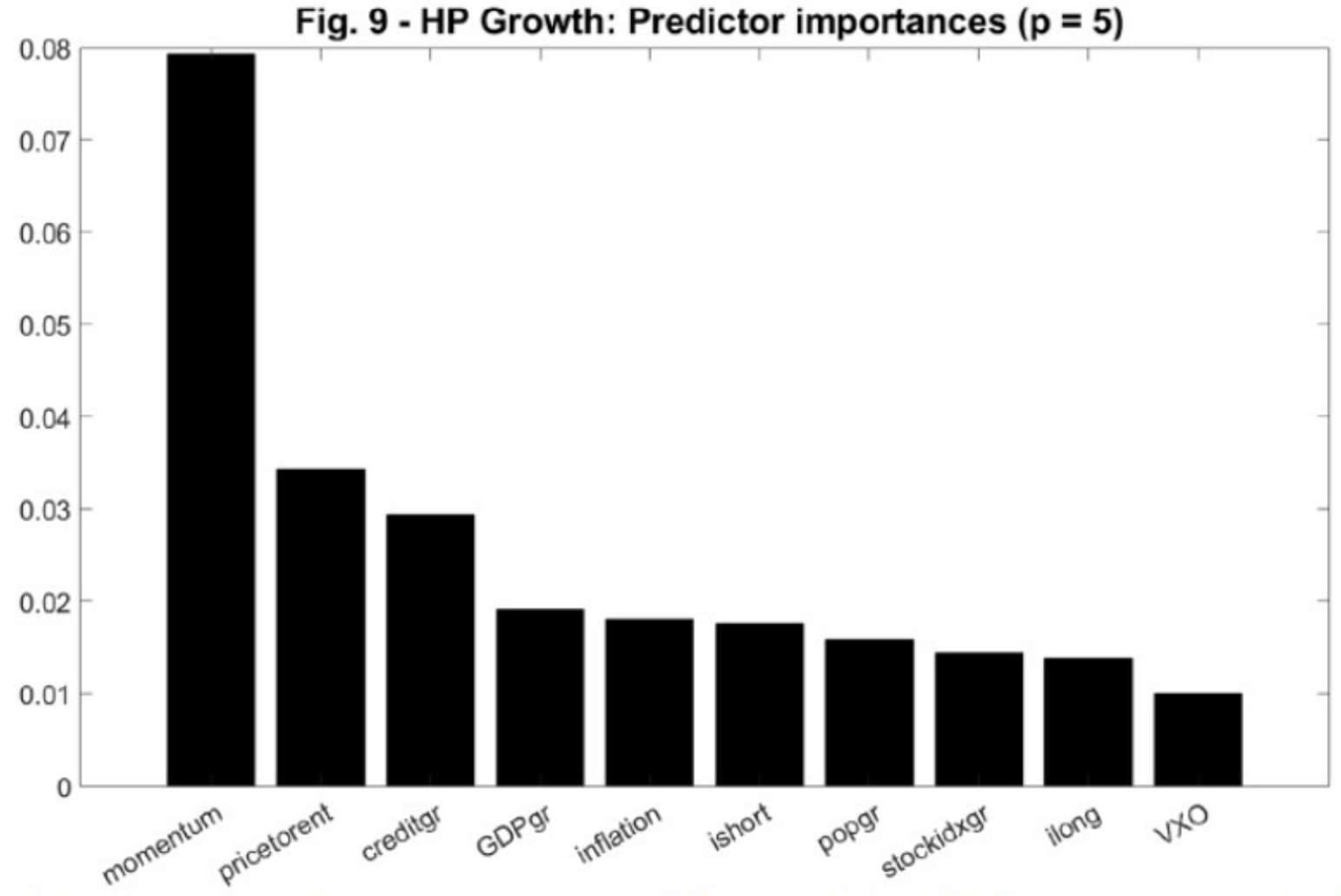

Mean absolute Shapley values (predictor importances) for each of the 10 features used in the baseline house price growth model when the minimum number of observations per parent node is set to five.

## 10 Concluding Remarks

All in all, this study has shown that a simple well-established machine learning model – random forests – can both, enhance our understanding of in-sample economic relations that prevail in housing markets, as well as considerably improve house price forecasts. Root mean square errors (RMSE) and mean absolute forecast errors (MAE) reductions exceed 40%. Importantly, these very large gains in predictive accuracy were obtained without relying on either big or granular data.

The bounded complexity of the random forest model of the housing market presented here was instrumental to aid the economic interpretation of the results. House price momentum, initial valuation and credit growth emerged as the most important determinants of prices – with little variation across countries. Further, partial effects suggest that house price growth is maximized when CPI inflation is in the 0–3% range, declining as the pace of overall price increases surpasses this level. Housing is found to be a rather poor hedge against high levels of inflation.[8] Importantly, the derivation of partial effects shows that the random forest model is not a black box.

While price momentum does stand out as a key driver of following year house price dynamics, economic fundamentals taken together appear to explain the bulk of price dynamics across countries and time. What this suggests is that government policies – especially those that affect household credit and, increasingly so, the yield curve - may be key for developments in housing markets and financial stability.

Future research may want to test if the above results extend to country specific studies based on more granular data. The role of country-specific characteristics of mortgage markets may also deserve further exploration.

[8] This is in line with Bekaert and Wang (2010) who report that, in advanced economies, real estate returns are negatively affected by inflation irrespective of whether a horizon of 1-, 3- or 5 years is used.

## Technical Appendix

**Data Sources:**

House price growth: Yearly growth rate of nominal house price index, computed from data obtained from the Bank for International Settlements;

Credit growth: Annual growth rate of nominal household credit, computed from data obtained from the Bank for International Settlements;

Inflation, GDP growth, stock index and population growth: The World Bank, Development Indicators database.

Short and long-term interest rate, price to rent index: OECD.

## Table A1 - Summary Statistics

| | mean | std. dev. | min | max |
|---|---|---|---|---|
| house price gr | 4.62 | 6.40 | -16.96 | 27.69 |
| inflation | 2.10 | 1.72 | -1.33 | 15.74 |
| GDP growth | 2.42 | 2.00 | -5.70 | 8.42 |
| short int. rate | 3.72 | 3.50 | -0.78 | 21.11 |
| long int. rate | 4.67 | 2.91 | -0.36 | 15.71 |
| stock index gr | 6.78 | 17.05 | -39.08 | 66.67 |
| credit gr | 6.06 | 4.71 | -3.49 | 30.13 |
| pop growth | 0.64 | 0.44 | -1.85 | 2.25 |
| price / rent ratio | 87.14 | 27.98 | 33.57 | 185.27 |
| VXO | 20.16 | 7.63 | 9.46 | 40.40 |
| house price growth by country | | | | |
| BEL | 5.51 | 3.60 | 0.13 | 16.14 |
| CAN | 5.21 | 5.76 | -7.13 | 21.87 |
| CHE | 1.69 | 4.96 | -6.33 | 21.96 |
| DEU | 2.15 | 2.89 | -1.74 | 8.64 |
| DNK | 4.07 | 6.83 | -10.40 | 22.66 |
| GBR | 6.01 | 7.71 | -8.95 | 27.69 |
| JPN | -2.04 | 6.03 | -14.86 | 9.62 |
| KOR | 3.44 | 6.53 | -12.29 | 21.15 |
| NLD | 5.49 | 5.67 | -4.97 | 20.10 |
| NOR | 5.49 | 7.05 | -13.65 | 15.14 |
| NZL | 6.37 | 6.45 | -8.93 | 24.36 |
| SWE | 5.90 | 6.56 | -15.28 | 19.12 |
| USA | 4.03 | 6.52 | -16.96 | 15.34 |

## Table A2 – OLS benchmarks

| D.V.: nominal house price growth | | |
|---|---|---|
| | AR(1) | OLS |
| constant | 1.673*** | 9.66*** |
| | 5.460 | 6.004 |
| lagged house price growth | 0.575*** | 0.427*** |
| | 11.64 | 8.18 |
| lagged inflation | | -0.085 |
| | | 0.22 |
| lagged GDP growth | | 0.201 |
| | | 0.15 |
| lagged short-term rate | | -0.472** |
| | | 2.27 |
| lagged long-term rate | | -0.025 |
| | | 0.10 |
| lagged stock market gain | | -0.021 |
| | | 1.24 |
| lagged credit growth | | 0.213*** |
| | | 2.80 |
| lagged pop. Growth | | -0.070 |
| | | 0.12 |
| lagged price/rent ratio | | -0.077*** |
| | | 6.94 |
| lagged VXO | | -0.004 |
| | | 0.11 |
| observations | 374 | 374 |
| R2 | 0.355 | 0.455 |
| Adjusted R2 | 0.353 | 0.440 |
| F-statistic | 204.0*** | 30.2*** |

Note: *t-statistics* based on robust standard errors. ***/**/* denote statistical significance at 1/5/10% confidence level.

**Figure A1 – Nominal House Prices over Time**
(base: 2010 = 100)

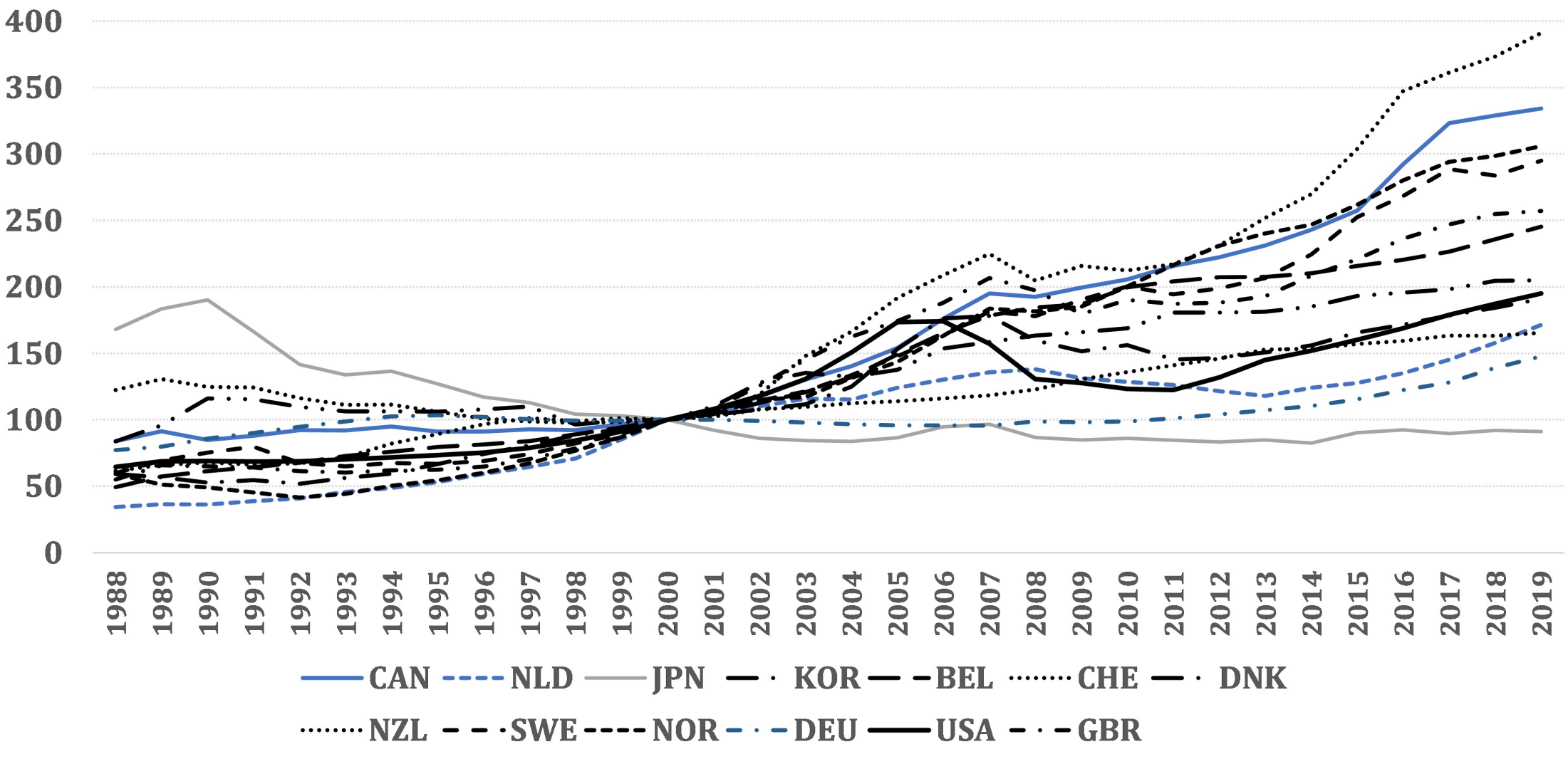

## Fig. A2 – Example of Grown Regression Tree (only top)

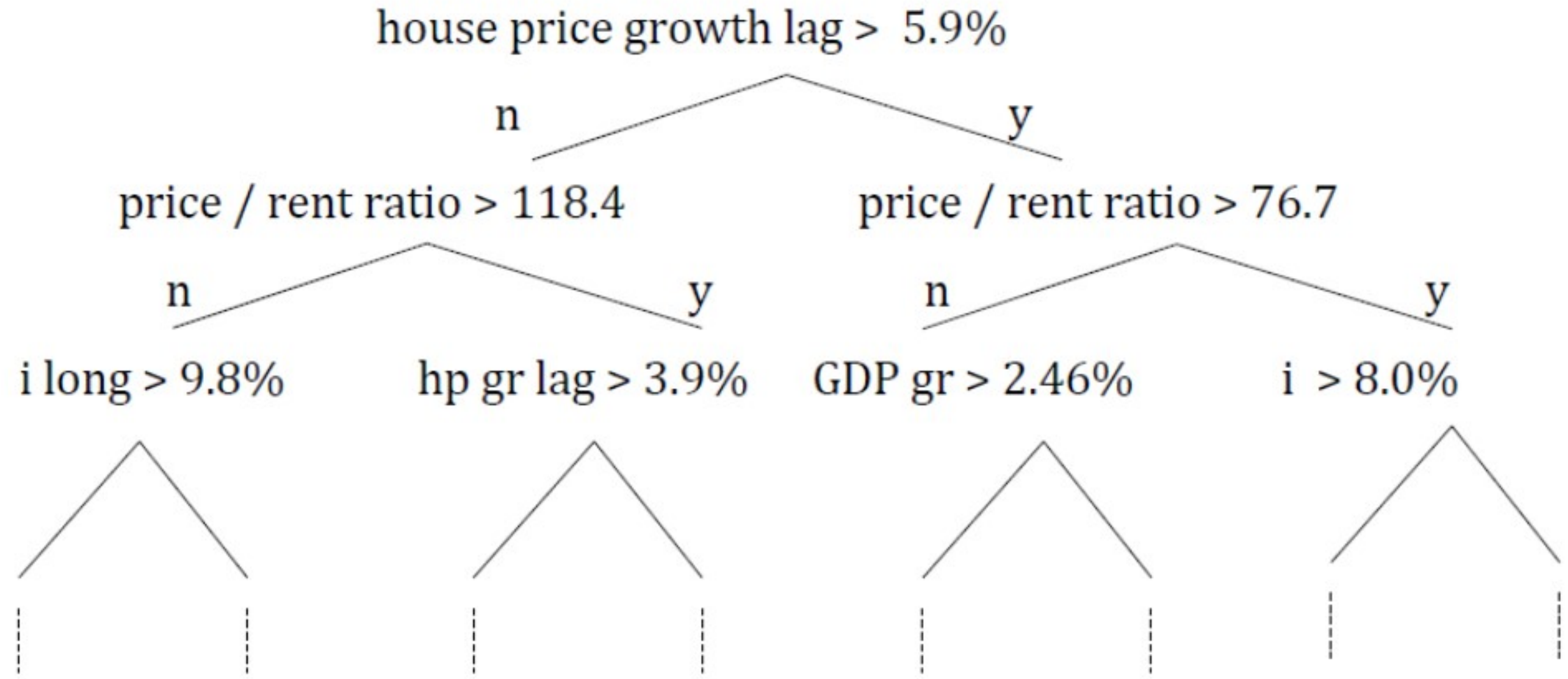

This figure shows the top of a regression tree created by the algorithm, with the respective splitting criteria.